\documentclass{article}

\usepackage{arxiv}

\usepackage[utf8]{inputenc} 
\usepackage[T1]{fontenc}    
\usepackage{hyperref}       
\usepackage{url}            
\usepackage{booktabs}       
\usepackage{amsfonts}       
\usepackage{nicefrac}       
\usepackage{microtype}      
\usepackage{cleveref}       
\usepackage{lipsum}         
\usepackage{graphicx}

\usepackage{natbib}
\usepackage{doi}
\usepackage{amsthm}
\usepackage{bmpsize}

\title{Practical Skills Demand Forecasting via Representation Learning of Temporal Dynamics}

\author{ \href{https://orcid.org/0000-0002-9365-9378}{\hspace{1mm}Maysa M. Garcia de Macedo} \\
	IBM Research\\
	Sao Paulo, Brazil\\
	\texttt{mmacedo@bbr.ibm.com} \\
	\And
	\href{https://orcid.org/0000-0003-1827-0554}{\hspace{1mm}Wyatt Clarke} \\
	IBM Research\\
	Yorktown Heights, Ny, United States\\
	\texttt{wyatt.clarke@ibm.com} \\
	\And
	 Eli Lucherini \\
	 Princeton University \\
	 Princeton, NJ, United States \\
	 \texttt{el24@cs.princeton.edu} \\
	 \And
	 Tyler Baldwin \\
	 IBM Research \\
	 San Jose, California, United States \\
	 \texttt{tbaldwin@us.ibm.com} \\
	 \And
	 Dilermando Queiroz Neto\\
	 IBM Research \\
	 Sao Paulo, Brazil \\
	 \texttt{dilermando.queiroz@ibm.com} \\
	 \And
	 Rogerio de Paula \\
	 IBM Research\\
	 Sao Paulo, Brazil \\
	 \texttt{ropaula@br.ibm.com} \\
	 \And
	 \href{https://orcid.org/0000-0002-7610-2738}{\hspace{1mm}
	 Subhro Das} \\
	 MIT-IBM Watson AI Lab, IBM Researc \\
	 Cambridge, MA, United States \\
	 \texttt{subhro.das@ibm.com} \\
}

\renewcommand{\shorttitle}{Practical Skills Demand Forecasting}

\hypersetup{
pdftitle={A template for the arxiv style},
pdfsubject={q-bio.NC, q-bio.QM},
pdfauthor={David S.~Hippocampus, Elias D.~Striatum},
pdfkeywords={First keyword, Second keyword, More},
}

\begin{document}
\maketitle

\begin{abstract}
  Rapid technological innovation threatens to leave much of the global workforce behind. Today’s economy juxtaposes white-hot demand for skilled labor against stagnant employment prospects for workers unprepared to participate in a digital economy. It is a moment of peril and opportunity for every country, with outcomes measured in long-term capital allocation and the life satisfaction of billions of workers. To meet the moment, governments and markets must find ways to quicken the rate at which the supply of skills reacts to changes in demand. More fully and quickly understanding labor market intelligence is one route.
In this work, we explore the utility of time series forecasts to enhance the value of skill demand data gathered from online job advertisements. This paper presents a pipeline which makes one-shot multi-step forecasts into the future using a decade of monthly skill demand observations based on a set of recurrent neural network methods. We compare the performance of a multivariate model versus a univariate one, analyze how correlation between skills can influence multivariate model results, and present predictions of demand for a selection of skills practiced by workers in the information technology industry.

\end{abstract}

\newtheorem{remark}{Remark}


\section{Introduction}
As the nature of work changes in the face of emerging technologies, students and workers must face the reality of a skills gap. This gap between the skills possessed by workers and those demanded by employers could leave millions of positions unfilled in the next 10 years, with important economic and human implications.  
Accurate predictions of future labor market conditions can help today's learners maximize their future productivity and meet employers' needs. While only individuals -- students and workers -- own it, many institutions cultivate and curate human capital. Schools, governments, and companies can all make better human capital decisions when armed with accurate labor market predictions.
When decision makers -- students, schools, governments, companies -- appeal to recent trends to guide their decision making, they are making implicit forecasts about future trends. In this paper, we explore what value can be added by using the same recent trends to form explicit forecasts with machine learning.

The contribution of this paper is to demonstrate how well past skill demands contained in online job advertisements can be used to predict future skill demands. Specifically, we make the following contributions:
\begin{enumerate}
    \item A pipeline with recurrent neural network (RNN) architecture that can flexibly switch between multivariate and univariate models. 
    \item Applying RNN to skills data. The nonlinearity of RNNs allows these models to perform well on multivariate skills demand prediction. 
    \item Demonstration of the tradeoffs between multi vs univariate modeling of skills demand, which is important given the potentially large number of skills to be forecast.
\end{enumerate}

In recent years, there has been increased focus on skills utilized in the labor market, understood as an unbundling of occupations. For example, the T3 Innovation Network is an open source community working to ``create more equitable and effective learning and career pathways.'' The T3 Innovation Network envisions the adoption of digital wallets with verifiable, standardized skill credentials for workers -- elevating ``competencies'' above occupations.\footnote{https://www.uschamberfoundation.org/t3-innovation}

Our work supports one of the goals of the T3 Innovation Network, to ``Analyze, compare, and translate competencies within and across industries using artificial intelligence and machine learning''~\cite{T3OnePage}. Additionally, our approach of directly forecasting skills demand using online job postings has been recognized by the Labor Market Information Council as a crucial nascent research direction to help workers understand future skill demands \cite{Bonen2021}.

The remainder of this paper is structured as follows. Section~\ref{sec:related} lists the prior art and Section~\ref{sec:problem} describes the data used for this work. Section~\ref{sec:method} describes the proposed method for predicting skills demand from online job posting data. The results are presented and discussed in Section~\ref{sec:result} along with their practical relevance, finishing with the paper conclusion in Section~\ref{sec:conclusion}.

\section{Background and related work}
\label{sec:related}

The value of up-to-date labor market data has long been recognized and studied. Since 1938, the U.S. Department of Labor has surveyed domain experts to provide detailed labor market analysis, starting with the Dictionary of Occupational Titles (DOT) and continuing with today's Occupational Information Network (O*NET).

This shift in focus toward skills has been enabled by innovations in natural language processing (NLP) and the digitization of labor market-related text. Private companies now gather and analyze large online text corpuses to glean and sell labor market intelligence. Job ads provide valuable details about labor demand, while r\'{e}sum\'{e}s and social media profiles illustrate the supply of different types of workers. Also, the course curriculum provides information about the skills imparted upon students at colleges or other training institutions who will soon enter into the labor market~\cite{yu-2021acl}. 

Increasingly, academic researchers and even the press use this privately curated labor market text data to characterize the labor market \cite{Deming2018,Porter2021}.

Government surveys can establish the typical skills used in an occupation and the number of workers who practice an occupation -- but they rarely directly connect workers to skills. The innovation of text data is observing individual jobs demanding individual tasks to be performed (or individual workers supplying specific skills). This new granularity has supported the use of economic models that, again, highlight skills as the primary unit of analysis \cite{Autor2003} 

One advantage of a skills-first view of the labor market is that it better captures the impact of recent technological changes on labor demand than earlier economic models have done \cite{Acemoglu2011}. Many authors have written about the huge potential for artificial intelligence (AI) to transform work -- sometimes in cataclysmic terms \cite{Frey2017,BrynjolfssonMcAfee2014,Acemoglu2019,Gathmann2010}.\footnote{\cite{Adler1992} summarizes America's modern history of anxiety around humans' role in work since 1950. \cite{Autor2015} gives an accessible analysis of why humans still have jobs after 200 years of technology-driven automation.} While there is uncertainty about the degree of change in the offing, most writers agree that it will be best characterized at the skill/task level \cite{BeaudryGreenSand2015,Bartel2007,fleming2019future}.

Which occupations become more or less common is hard to guess, but commentators are in near unison that AI will eventually change the task content of most jobs.

\subsection{Forecasting with deep learning models}
In finance and economics, time series forecasting is traditionally performed with ARIMA models.

However, ARIMA models do not capture nonlinear couplings between variables, which can severely limit their prediction accuracy.

More recently, deep learning models, such as long short-term memory (LSTM)~\cite{Hochreiter} and Gated Recurrent Unit (GRU), have gained popularity in the domain \cite{krauss2017deep,fischer2018deep,yamak2019comparison}. 

These models have shown superior predictive performance when applied to time series presenting high volatility in various use cases such as the prediction of monthly financial time series \cite{siami2018comparison}, gold prices \cite{livieris2020cnn}, and demand \cite{abbasimehr2020optimized}. 
In this paper, we build a pipeline and compare the performance of several deep learning models at predicting skill demand.

\subsection{Job skill demand forecasting}

Das \emph{et al} \cite{das2020aies} introduced the paradigm for quantifying skill demand and made single-step 1-month-ahead forecasts of skill ``clusters'' using an Autoregressive Integrated Moving Average (ARIMA) model. In this work, we are addressing skill demand forecasts for individual skills -- a level we hope is useful to aid individual workers' choices -- 12 months into the future. We also move beyond ARIMA, comparing the performance of several neural network architectures. Overall, our goal is to demonstrate actionable predictions to guide workers' and institutions' human capital investment decisions.

\section{Problem Statement \& Data}
\label{sec:problem}

\begin{figure*}[!t]
    \centering
    \begin{tabular}{ccc}
      \includegraphics[width=0.3\linewidth]{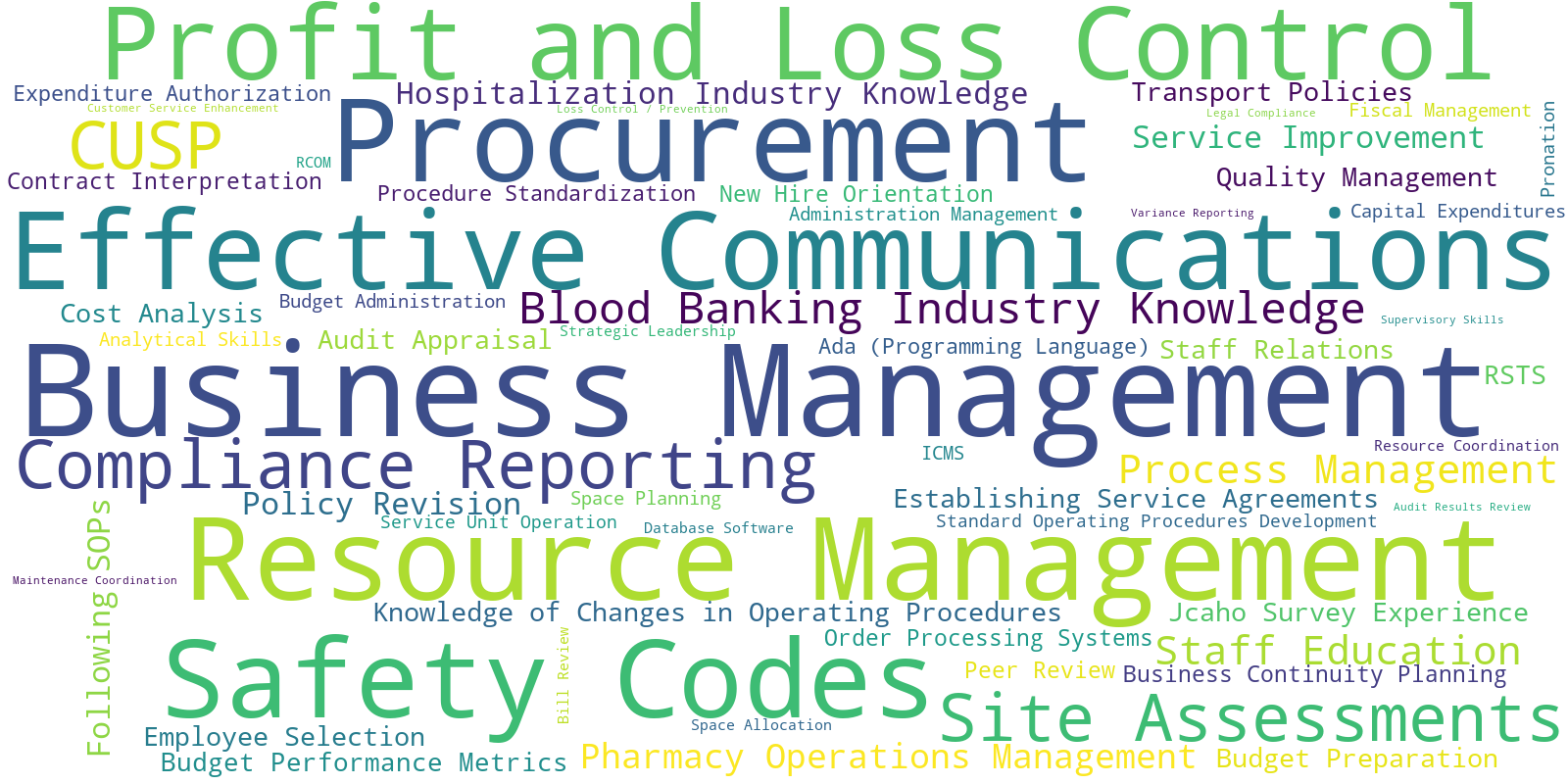}    & 
      \includegraphics[width=0.3\linewidth]{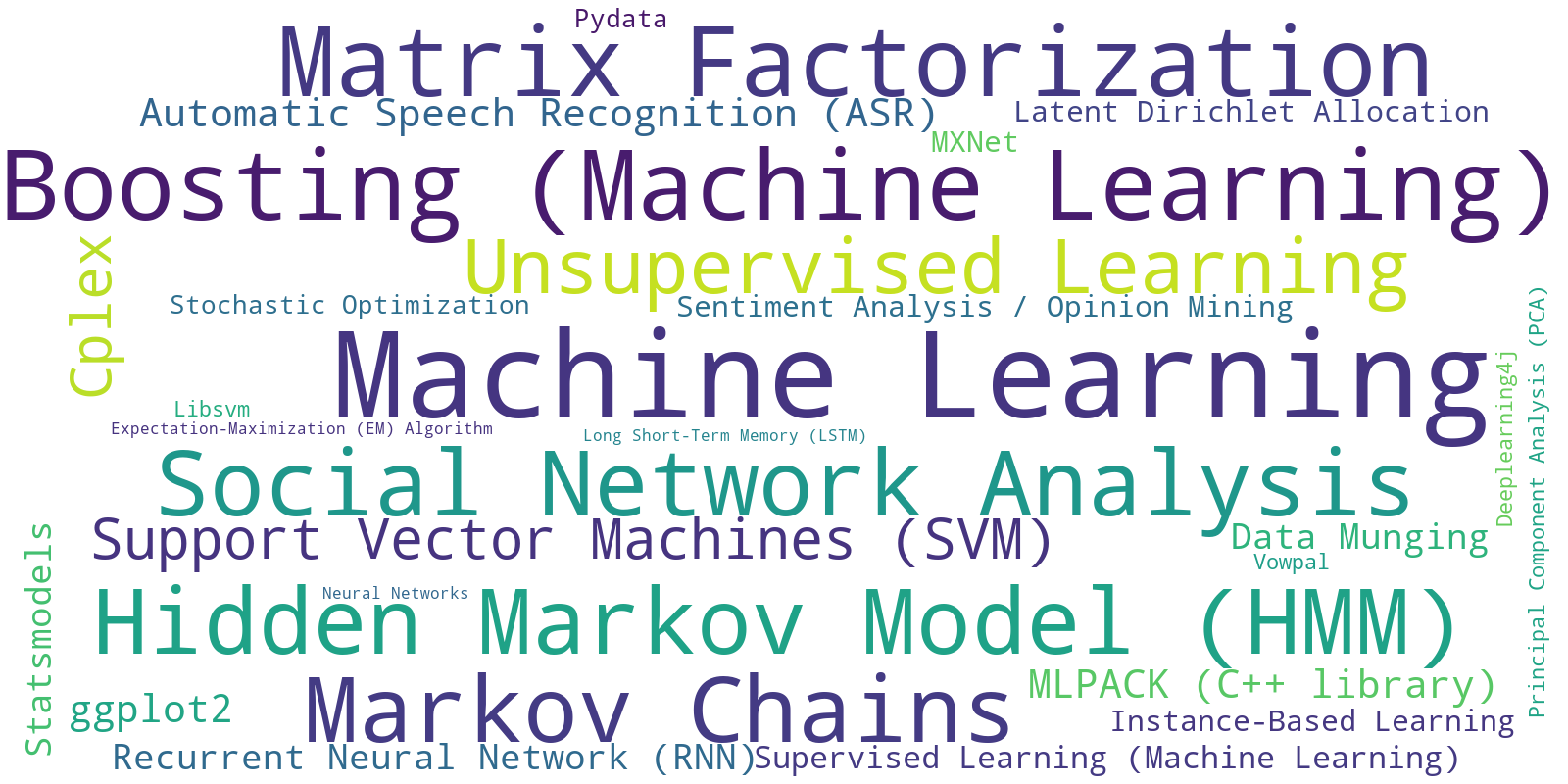}   &
      \includegraphics[width=0.3\linewidth]{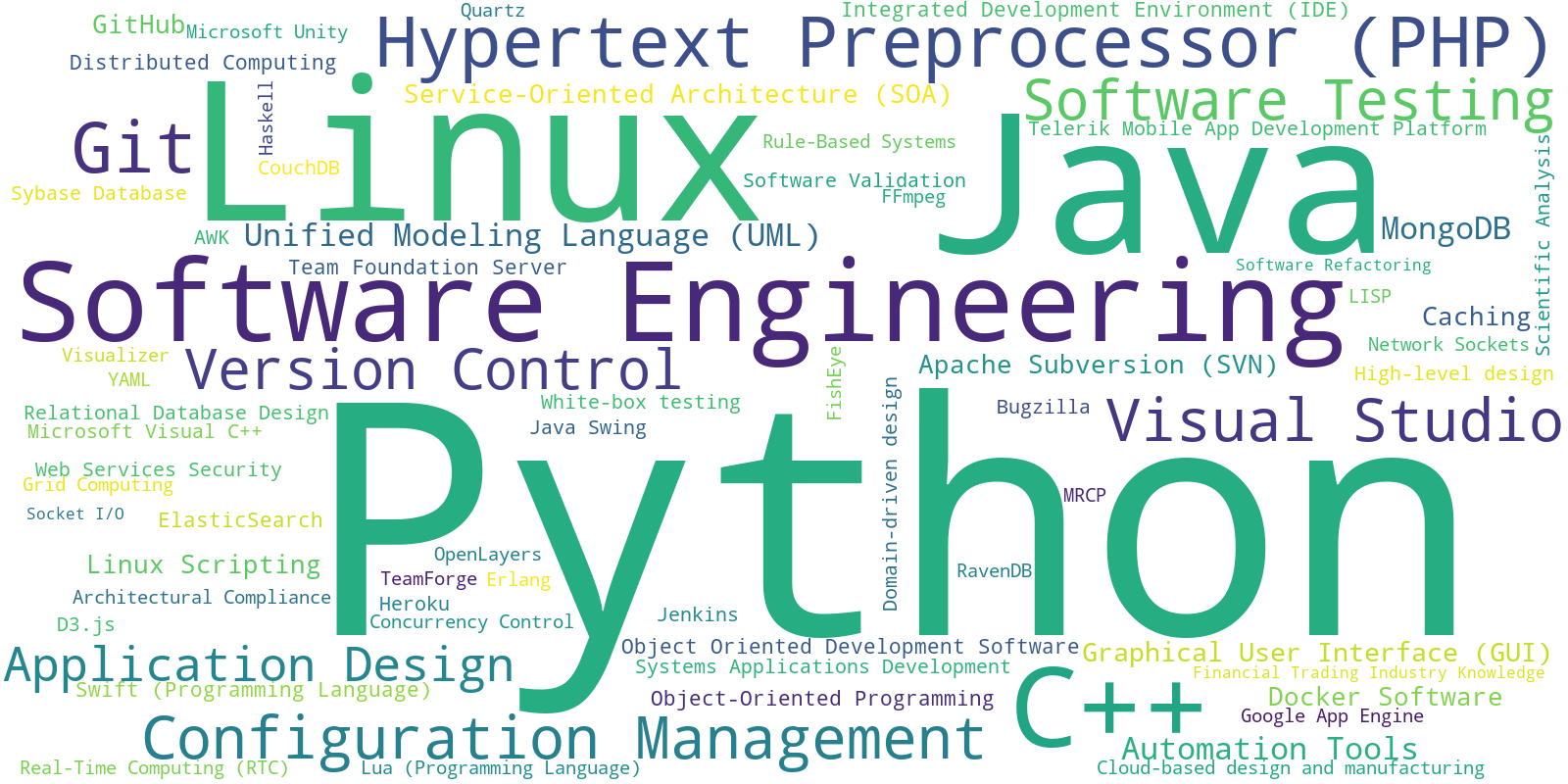}   \\ 
     (a)  & 
      (b)  &
      (c)  \\ 
    \end{tabular}
     \caption{Word cloud representing all skills at the same cluster. Each word cloud represents skills with high similarity with the key skill: (a) `Business Management', (b) `Machine Learning' and (c) `Python'.}
    \label{fig:wordcloud}
\end{figure*}

The remainder of this paper presents and discusses time series forecasts of the frequency with which certain words or phrases appear in online job advertisements, indicating the relative demand for skills that those phrases represent. The data provider, Emsi Burning Glass (EBG), scrapes text from internet websites, parses the text into individual job ads, and applies a proprietary entity extraction pipeline to each advertisement. Metadata including occupation and industry are assigned to each ad, along with indicators for the appearance of roughly 17 thousand concepts that Emsi Burning Glass describes as ``skills.'' It is these structured data resulting from the entity extraction pipeline that we work with.

Online job ads are unrivaled as a source of information at scale about the requirements for jobs. Previous authors have shown that job ads also correlate with relevant labor market time series including job openings, hiring and employment \cite{Deming2018}. However, job ads have known limitations, including that the ratio of advertisements to jobs varies between occupations, industries, locations and many other axes. They are often best used in combination with representative surveys \cite{Carnevale2014}.\footnote{For example, the Conference Board maintains an index of trends in online job advertising called the Help-Wanted OnLine (HWOL). It combines data from the U.S. Bureau of Labor Statistics' (BLS) Job Openings and Labor Turnover Survey (JOLTS) and job ads from Emsi Burning Glass. JOLTS data are broken out for only high level industries and geographies. HWOL uses JOLTS data as a skeleton and interpolates more detailed estimates for industries, occupations, and geographies using EBG data \cite{ConferenceBoard2019}.}

Accordingly, Das \emph {et al}~\cite{das2020aies} develop a concept of \textit{skill shares} that will be the subject of this paper's forecasts.\footnote{\cite{das2020aies} actually use the term ``task-share'' but note the interchangeability of the terms ``skill'' and ``task'' in this context. We prefer to use ``skill'' since the focus is on workers' human capital.} Skill shares combine EBG job ad data with U.S. Bureau of Labor Statistics (BLS) occupational employment statistics to approximate the percentage of workers in an occupation practicing a set of skills. Skill share is used as a proxy for the demand from employers for workers who can perform a skill. 
It is calculated using the share of job ads within an occupation that mention the specified skill(s) in a given month, multiplied by the share of U.S. workers employed in that occupation in that month, summed over occupations.\footnote{This calculation of skill share assumes the skills mentioned in job ads are representative of the skills used by already-employed workers in the same job, which is why we characterize it as an approximation.} Equation 1 illustrates a skill share calculation for skill $i$ within a single occupation $j$ during month $t$. Note that skill share is additive, meaning aggregated skill shares for multiple skills or occupations can be obtained by adding the individual figures.\footnote{We also tested an alternative specification of skill share where the occupation's employment share $\frac{\textrm{emp}_{j,t}}{\textrm{all emp}_{t}}$ is held constant at $t=2010$ levels. Doing so isolates the change in skills usage attributable to within-occupation changes rather than to a rebalancing of occupational employment in the economy. In most cases, the time series is barely altered.}

\begin{equation}
\textrm{skill share}_{i,j,t} = \frac{\textrm{ads}_{i,j,t}}{\textrm{all ads}_{j,t}} \times \frac{\textrm{emp}_{j,t}}{\textrm{all emp}_{t}}
\end{equation}

For the skill share calculations used in this paper, occupations are specified using Standard Occupation Classification (SOC) 2010 six-digit codes. This most detailed level of SOC has roughly eight hundred occupations. Additionally, we calculate using the most granular level of EBG skill data, with roughly 17 thousand named skills. To get monthly observations, we interpolate the annual BLS employment statistics using a piecewise linear function.
We chose such a highly detailed view of skill shares so that our forecasts can be relevant to individual workers. Given that skill share is additive, we can also aggregate skill shares to any desired set of skill-occupation combinations. For tractability, we selected a few interesting data series to forecast. 

\subsection{Scope of the forecasting framework}

Next we make a few remarks to discuss some pitfalls we sought to avoid, and thus our criteria for which forecasts to present. These include the impact of COVID-19, the length of time it takes to acquire skills, how long they are relevant, and which skill-occupation combinations are well represented in online job ads. 
\begin{remark}
The EBG data used for this project extend from 2010 to 2020. During 2020, the COVID-19 pandemic caused unprecedented swings in employment levels for many occupations, which is not well represented by our monthly interpolation of annual BLS employment statistics. Thus, we omit 2020 and calculate a panel of skill shares with 120 monthly observations from 2010 to 2019 for each skill-occupation combination. 
\end{remark}

\begin{remark}
The horizon over which we can make credible forecasts (about a year) is much shorter than people's learning and working lives -- up to 20 years of schooling followed by around 40 years of working. We cannot say much about the evolving value over a lifetime of foundational skills like basic mathematics or grammar. Instead, our work should focus on granular, relatively short-term human capital investments, in areas where workers understand that the skill demands (often technologies) change fast. Deming \emph{et al} ~\cite{Deming2018a} show that STEM careers involve skills whose value decay relatively quickly.
\end{remark}
\begin{remark}
We want to pick occupations and skills that are well represented in job postings. Certain skills that are useful in a job might be left implicit in job ads (e.g., Microsoft Excel for a data scientist), so we want to focus on skills that would be salient in a given occupation \cite{Bonen2021}. Also, while professional jobs have been almost universally advertised online during the past decade, many jobs that require less education continue to be advertised offline through physical signs or word of mouth \cite{Carnevale2014, LMIC2020}. Skill demands in those occupations are of course impacted by technology and subject to change, but their incomplete representation creates a selection problem: it is hard to know how the task content of jobs advertised online differs from that of the jobs advertised offline. 
\end{remark}
Bearing these considerations in mind (and the makeup of our likely readership), all the forecasts we present are for skills practiced by a subset of SOC occupations, Computer and Mathematical Occupations.\footnote{\url{https://www.bls.gov/soc/2010/2010\_major\_groups.htm\#15-0000}} We employ a methodology to group skills into a few clusters by computing similarity between skills, detailed in the next subsection. These subsets of clustered data are suitable to measure, for example, the increase in demand in the labor market for workers who know Python (skill).

\subsection{Skill clusters based on skill similarity}

To choose groups of skills to study together, we utilize the word2vec embedding method~\cite{mikolov_efficient_2013} from natural language processing.  Word2vec uses co-occurrence of words or phrases with one another in text to learn an embedding of these terms in a lower dimensional space where distance in the space is indicative of similarity between terms.  Training was done using skill lists associated with each job posting from 10 years of Emsi Burning Glass data, from 2010 to 2019.  The skipgram training method was utilized with an embedding size of 30.  Cosine similarity was used to determine the most similar skills, which were then clustered via K-means clustering.

In summary, we first identify 11 different key skills, then for each key skill we build a cluster of skills around it using the skill similarity technique described above. As a result, in each of these cluster datasets we have a key skill and other adjacent skills that have a strong similarity to each other. Figure \ref{fig:wordcloud} shows three different word clouds which depict the skills in three different clusters. On average, there are 52 skills per dataset.
The key skills are:
`Agile development',
`Business management',
`Cloud computing',
`Cryptography',
`Data science',
`Decision support',
`Machine learning',
`Python',
`Precision Agriculture',
`Security assessments', and
`Quantum computing' / `natural language processing' (in the same cluster).

In addition to these eleven skill cluster datasets, we consider an additional dataset constructed differently than the rest. It is composed of aggregated groups of skills that EBG calls \textit{Skill Cluster Families}~\cite{burning2019mapping}, which map roughly to career areas (e.g. Information Technology, Finance, and Health care). We include this more aggregated dataset to allow direct comparison to past studies that consider skill demand forecasting.

\section{Methods}
\label{sec:method}
\begin{figure}[t!]
    \centering
    \includegraphics[width=\linewidth]{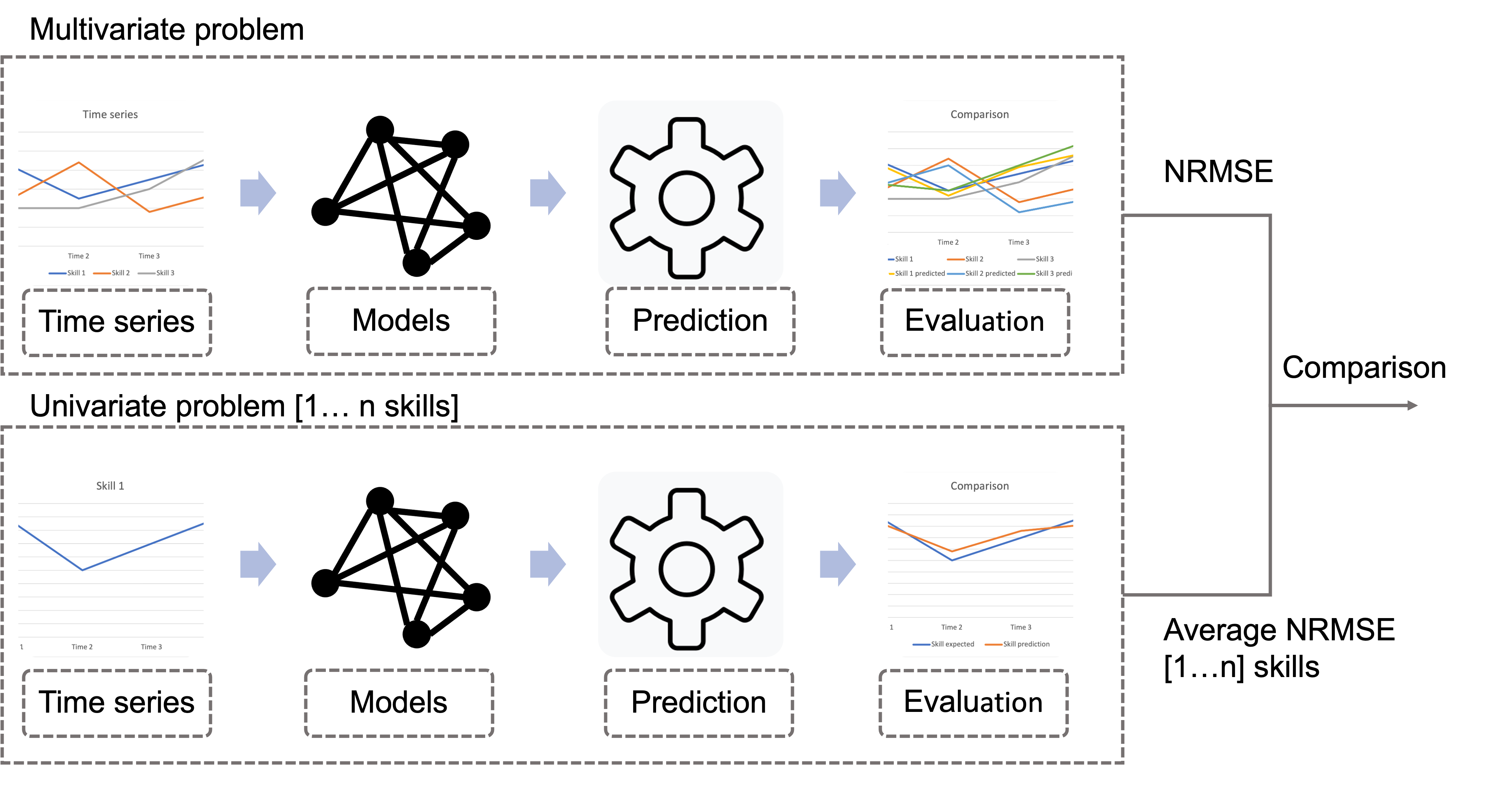}
    \caption{Scheme of the pipeline that uses recurrent neural networks to predict skills demand. We look for the best result with the multivariate model and compare it with the average of the best results for each skill in the univariate model.}
    \label{fig:illustration}
\end{figure}

The primary objective of this endeavor is to develop a skills demand prediction pipeline that receives monthly skill share values for the above set of skills and makes a prediction for the next 6, 12, 24 and 36 months. This prediction can be performed basically in two ways: univariate, when we predict a single time series and multivariate, when we predict 2 series or more at the same time. Regarding the months ahead to be predicted, we can also differentiate the models in two ways, the first is when the model predicts one month per run and the errors accumulate for each predicted month, and the second is when \emph{x} months are predicted at a single shot. For this work, we tested both multivariate and univariate forms and only the one-shot prediction for the months. As there are several state-of-the-art prediction methods, we trained three state-of-art recurrent neural network methods on the skill-share datasets: long short-term memory (LSTM)~\cite{10.1162/neco.1997.9.8.1735}, convolutional neural networks combined with LSTM (CNN+LSTM)~\cite{7558228} and gated recurrent units (GRU)~\cite{Cho2014LearningPR}. Figure \ref{fig:illustration} graphically describes the pipeline employed in this paper. In this section, we detail the pre-processing techniques for the skills time series and the experiment design for performing demand forecasts utilizing these methods.

\subsection{Time series}
The \textit{skill shares} time series data, derived from online job postings, ranges from 2010 to 2019, aggregated at a monthly frequency. Note that the time series could have been constructed at a higher frequency (weekly or daily), however higher frequency aggregations tend to incorporate noise and make the data sparser. Hence, we considered 120 months of observations for each series, where the first nine years of skill shares are used as training data (from 2010 to 2018) and the final year as testing data (2019), in case of predicting 12 months ahead.

Even at a monthly frequency the time series exhibits a significant degree of fluctuation which necessitates standard pre-processing operations. Firstly, we perform a smoothing average with a three-month moving window to filter out noisy fluctuations in the time series signals; secondly, this smoothed signal is \emph{first-differenced} to best capture changes in the dynamics by removing the trend in the signals. The resultant time series data is fed as an input sequence for the recurrent neural networks in a format consisting of $n$ time-point lags, where $n$ is the number of previous months of the time series being fit to generate the forecast. Finally, we standardize the skills by removing the mean and scaling to unit variance. This process is applied for training and testing separately.

\subsection{Models}

Deep learning models can capture nonlinearity in the evolution of skill shares over time, resulting in more accurate predictions. In this context, we train three different architectures, LSTM, CNN+LSTM and GRU to predict the dynamics of skill shares over time. We hypothesized that multivariate models are better suited to capture the correlations between skill shares than univariate models, so for this study a comparison is performed between multivariate and univariate models. However, multivariate deep learning models introduce two challenges. First, the models are computationally more complex and less likely to converge. The complexity is mainly driven by the number of time series that the model must predict at once. Second, the number of observations in our dataset is relatively small, making it harder to train deep learning models, which typically achieve higher accuracy with larger amounts of data.

As we desire to predict skill shares \emph{x} months ahead in one-shot, a modification in the last dense layer is performed for each of the three architectures. The size of the layer should be the number of months to be predicted times the number of skills to be predicted. Below we explain in more detail the variation of hyperparameters for each architecture. In all three architectures, we use Adam as the optimization algorithm with mean squared error (MSE) as the loss metric.

\subsubsection{LSTM}
This model refers to three layers of LSTM, each with a different size between 1 and 10 and with the number of neurons varying between 1 and 10, followed by a dense layer with the size of the number of months to be predicted multiplied by the number of skills to be predicted.

\subsubsection{CNN+LSTM}
A CNN+LSTM architecture involves using convolutional neural network (CNN) layers for attribute extraction from input data, combined with LSTM to support sequence prediction.
For this study, we used a one-dimensional CNN layer, where the kernel size varies between 2 and 64 and the number of filters varies from 1 to 10 (following the number of neurons of LSTM). We use strides equal to 1, ReLU activation, and same padding. This layer is followed by three different LSTM layers with size between 1 and 10 and the number of neurons varying between 1 and 10. Last, we have a dense layer with the size of the number of months to be predicted times the number of skills to be predicted. 

\subsubsection{GRU}
Similar to the LSTM model, the GRU model refers to three different-sized layers of GRU, where the number of neurons varies between 1 and 10, followed by a dense layer with the size of the number of months to be predicted times the number of skills to be predicted.

\begin{table*}[]
 \caption{Prediction of 12 months ahead of a multivariate model \emph{vs.} univariate model using the best parameters from multivariate model \emph{vs.} univariate model tuned for each skill}
    \label{tab:multi_v_uni}
\centering
\scalebox{0.92}{
    \begin{tabular}{lccccccccc   }
    \toprule

	&	\multicolumn{3}{c}{CNN+LSTM}			&	\multicolumn{3}{c}{LSTM}			&	\multicolumn{3}{c}{GRU}		\\	
\hline
DATASET	&	MULTI	&	UNI	&	UNI Tuned	&	MULTI	&	UNI	&	UNI Tuned	&	MULTI	&	UNI	&	UNI Tuned	\\
 \midrule
Agile Development	&	0.16	&	0.70	&	0.24	&	0.17	&	0.60	&	0.27	&	0.17	&	0.47	&	0.27	\\
Business Management 	&	0.08	&	0.32	&	0.13	&	0.09	&	0.32	&	0.14	&	0.08	&	0.48	&	0.14	\\
Cloud Computing     	&	0.42	&	0.64	&	0.25	&	0.42	&	0.49	&	0.28	&	0.44	&	0.50	&	0.28	\\
Cryptography        	&	0.30	&	0.62	&	0.30	&	0.32	&	0.66	&	0.34	&	0.28	&	1.03	&	0.33	\\
Data Science       	&	0.13	&	0.61	&	0.17	&	0.14	&	0.33	&	0.21	&	0.13	&	0.62	&	0.21	\\
Decision Support   	&	0.09	&	0.22	&	0.10	&	0.10	&	0.23	&	0.10	&	0.10	&	0.29	&	0.11	\\
Machine Learning   	&	0.40	&	0.87	&	0.38	&	0.44	&	0.76	&	0.44	&	0.42	&	0.97	&	0.42	\\
Precision Agriculture 	&	0.36	&	0.87	&	0.24	&	0.38	&	1.20	&	0.25	&	0.36	&	0.99	&	0.26	\\
Python 	&	0.13	&	0.50	&	0.22	&	0.13	&	0.79	&	0.24	&	0.14	&	0.44	&	0.24	\\
Quantum Computing   	&	0.37	&	0.35	&	0.18	&	0.38	&	0.32	&	0.20	&	0.38	&	0.40	&	0.20	\\
Security Assessments	&	0.29	&	0.55	&	0.26	&	0.28	&	0.63	&	0.28	&	0.30	&	0.49	&	0.28	\\
\hline
Average	&	0.248	&	0.568	&	0.225	&	0.258	&	0.575	&	0.250	&	0.254	&	0.607	&	0.249	\\
Standard Deviation	&	0.130	&	0.211	&	0.079	&	0.137	&	0.283	&	0.092	&	0.134	&	0.262	&	0.086	\\
\hline
\multicolumn{9}{l}{ Metric: Normalized Root Mean Square Error}\\
\bottomrule
    \end{tabular}
}   
\end{table*}

\subsection{Experiments}
\label{subsection:experiments}

We seek to answer a few questions by applying the models on the skills datasets. First and foremost, how well can these methods predict the demand for skills? What are the differences in performance and complexity of a multivariate model versus a univariate model? And, what are the best approaches to set hyperparameters? 

\begin{remark}
Our goal is to emphasize a practical approach to skills forecasting. EBG data tag over 17 thousand unique skills in job ads. An organization wanting to maintain forecasts for even a small share of those skills might be stuck with hundreds of univariate models. One way to simplify is to forecast multiple skills in a single multivariate model. While statistical and computational limits restrict the number of skills that can be modeled together, multivariate models could decrease the number of models requiring maintenance by at least an order of magnitude.
\end{remark}

Also, as noted before, we hypothesize that coupling (as measured by correlation) between variations in demand for different skills could be exploited by a multivariate model to produce more accurate predictions. 

Next, we explain the experiments we ran to probe these questions. Multiple values were tried for the number of month lags used in training, with $l \in$ $[12, 24, 36]$. Also, we varied the epochs, using the values $[50, 100, 500, 1000, 2000]$.

\subsubsection{Experiment: Multivariate}

This experiment comprises the training and testing of skill share predictions for each of the LSTM, CNN+LSTM and GRU recurrent neural networks and their subsequent comparison of results. Each dataset is trained considering all its skills and leaving only the months to be predicted aside for testing. The prediction is for 6, 12, 24 and 36 months ahead in one-shot. We perform hyperparameter tuning and display results of the best model.

\subsubsection{Experiment: Univariate using parameters chosen for the multivariate model}
In this experiment we compare the performance of the multivariate model with univariate models as follows: the set of parameters that provides the best result for the multivariate problem is also used in the univariate problem. If the multivariate problem has $n$ skills in its input data, a univariate experiment is run for each of the $n$ skills. At the end, the average normalized root mean square error (NRMSE) value of the $n$ univariate experiments is compared to the NRMSE value obtained in the multivariate experiment. 

\subsubsection{Experiment: Univariate using best parameters for each model}

For this experiment, we perform hyperparameter tuning to choose the best model for each univariate model related to the $n$ skills. In this way we make a comparison between the NRMSE value of the multivariate experiment and the average of the $n$ NRMSE values of the best univariate experiments. In this case we generate in total $n+1$ models, 1 multivariate and $n$ univariate.

\begin{table*}[]
 \caption{Prediction of 6,12,24 and 36 months ahead of a multivariate model, varying the RNN model}
    \label{tab:multi}
\centering
    \begin{tabular}{lcccccccccccc}
   \toprule
    &
    		\multicolumn{4}{c}{CNN+LSTM}&								\multicolumn{4}{c}{LSTM}&								\multicolumn{4}{c}{GRU}\\							
DATASET	&	6	&	12	&	24	&	36	&	6	&	12	&	24	&	36	&	6	&	12	&	24	&	36	\\
\midrule
Agile Development	&	0.12	&	0.16	&	0.41	&	0.50	&	0.12	&	0.17	&	0.45	&	0.51	&	0.12	&	0.17	&	0.52	&	0.50	\\
Business Management 	&	0.07	&	0.08	&	0.15	&	0.29	&	0.06	&	0.09	&	0.16	&	0.30	&	0.07	&	0.08	&	0.15	&	0.30	\\
Cloud Computing     	&	0.13	&	0.42	&	0.39	&	0.55	&	0.13	&	0.42	&	0.40	&	0.55	&	0.13	&	0.44	&	0.41	&	0.56	\\
Cryptography        	&	0.16	&	0.30	&	0.50	&	0.57	&	0.17	&	0.32	&	0.52	&	0.57	&	0.17	&	0.28	&	0.42	&	0.57	\\
Data Science       	&	0.11	&	0.13	&	0.45	&	0.31	&	0.11	&	0.14	&	0.44	&	0.33	&	0.13	&	0.13	&	0.44	&	0.32	\\
Decision Support   	&	0.06	&	0.09	&	0.14	&	0.14	&	0.06	&	0.10	&	0.14	&	0.14	&	0.06	&	0.10	&	0.12	&	0.15	\\
Machine Learning   	&	0.19	&	0.40	&	0.76	&	1.74	&	0.22	&	0.44	&	0.77	&	1.76	&	0.20	&	0.42	&	0.76	&	1.74	\\
Precision Agriculture 	&	0.32	&	0.36	&	0.62	&	0.60	&	0.35	&	0.38	&	0.65	&	0.62	&	0.35	&	0.36	&	0.59	&	0.60	\\
Python 	&	0.07	&	0.13	&	0.52	&	0.48	&	0.08	&	0.13	&	0.51	&	0.48	&	0.08	&	0.14	&	0.54	&	0.49	\\
Quantum Computing   	&	0.26	&	0.37	&	1.16	&	1.14	&	0.26	&	0.38	&	1.17	&	1.14	&	0.27	&	0.38	&	1.15	&	1.14	\\
Security Assessments	&	0.24	&	0.29	&	0.40	&	0.62	&	0.24&		0.28	&	0.41	&	0.69	&	0.24&		0.30	&	0.41	&	0.64	\\
\hline
\multicolumn{13}{l}{Metric: Normalized Root Mean Square Error}\\
																									
\bottomrule
    \end{tabular}
   
\end{table*}

\begin{figure}[t!]
    \centering
  \begin{tabular}{cc}
    \includegraphics[width=1\linewidth]{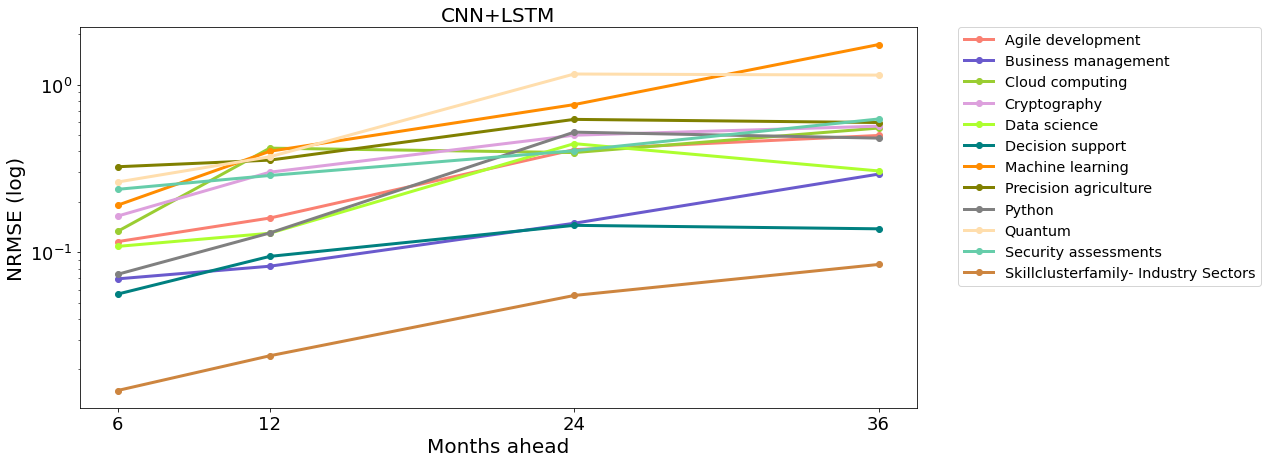}
    \end{tabular}
    \caption{Representation of the method performance along the months to be predicted. }
    \label{fig:monhs_ahead}
\end{figure}

\section{Results and Discussion}
\label{sec:result}

Results from the three experiments are displayed in Tables~\ref{tab:multi_v_uni} and \ref{tab:multi}. Table~\ref{tab:multi_v_uni} compares performance in a 12-months-ahead prediction of the multivariate experiment, univariate experiments each using the same parameters selected for the multivariate case, and univariate experiments with optimal parameters selected for each skill. Table~\ref{tab:multi} shows results of the multivariate model with different architectures and for different time horizons. 

We consider normalized root mean squared error (NRMSE) as the evaluation metric to compare the performance of the predictions with respect to the actual skillshare values. The NRMSE measure the prediction error of a one-shot, multi-step prediction of skill shares for the 6 to 36 months, which were kept aside as an evaluation sample. The comparison between the multivariate and univariate models is provided by the metric from the multivariate model with the average metric of all univariate results.

As demonstrated by Table~\ref{tab:multi_v_uni}, for the multivariate experiment CNN+LSTM performs best for four of the eleven datasets, with LSTM outperforming it for `Security assessment' and GRU showing the best performance for `Cryptography', and all the models perform similarly for the remaining datasets. On average, CNN+LSTM is the best architecture of the three experiments in 12-months-ahead forecasts. For the second experiment the univariate model was never the best; this is expected because the parameters are not personalized for each skill. The third experiment (univariate tuned) performed better than the first and second (multivariate and univariate with the same parameters). However an important limitation of the univariate tuned model is that is necessitates hyperparameter tuning for each skill. On average for multivariate experiments, the results of the CNN+LSTM model is very close to GRU. However, given its slightly lower standard deviation, CNN+LSTM seems like a better option than GRU.

\begin{figure*}[]
    \centering
    \begin{tabular}{cc}
      \includegraphics[width=0.49\linewidth]{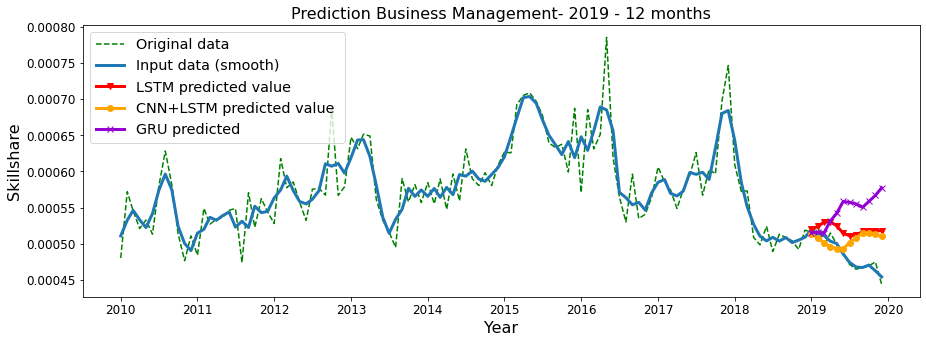}    & \includegraphics[width=0.49\linewidth]{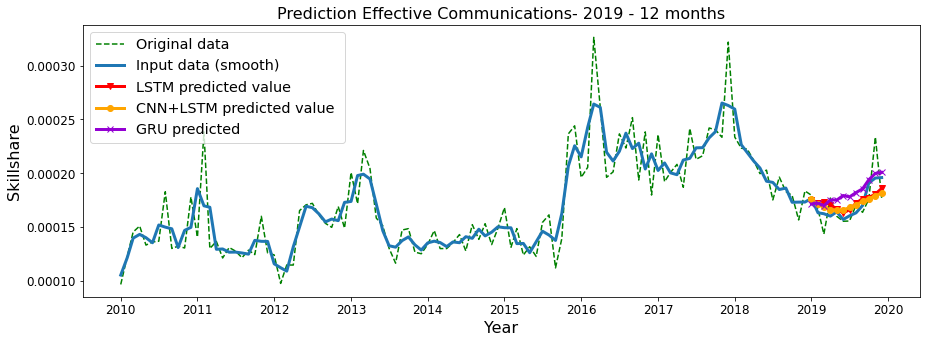}   \\
      \includegraphics[width=0.49\linewidth]{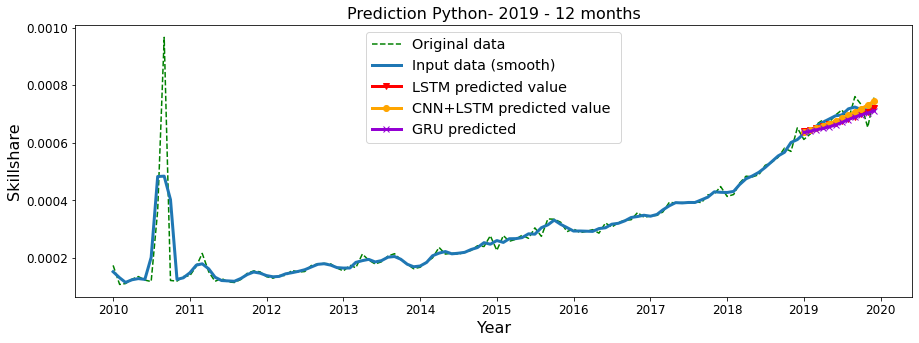}   & \includegraphics[width=0.49\linewidth]{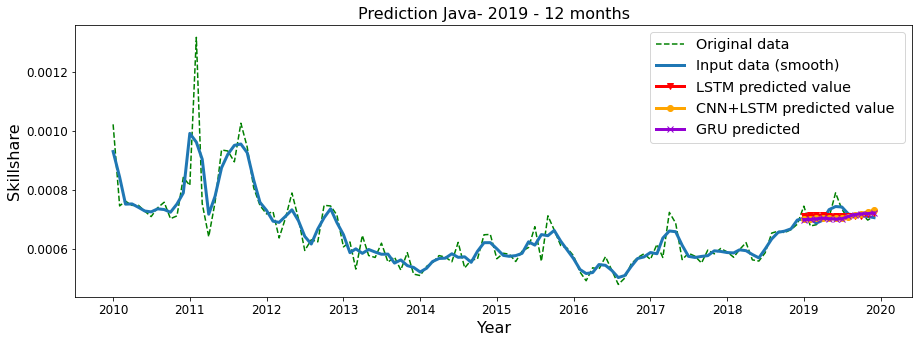}\\
      \includegraphics[width=0.49\linewidth]{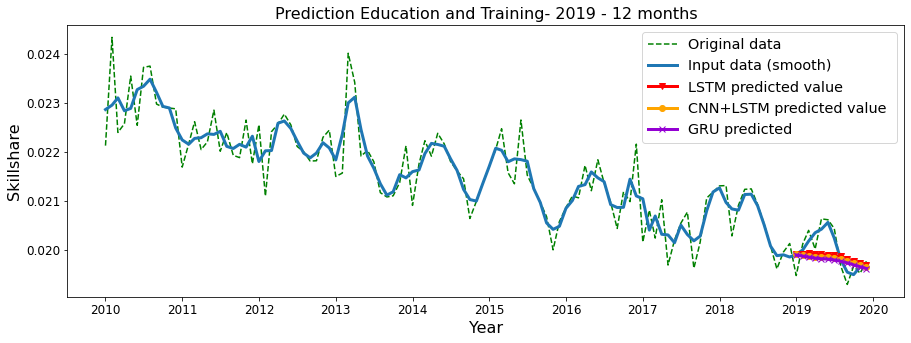}   & \includegraphics[width=0.49\linewidth]{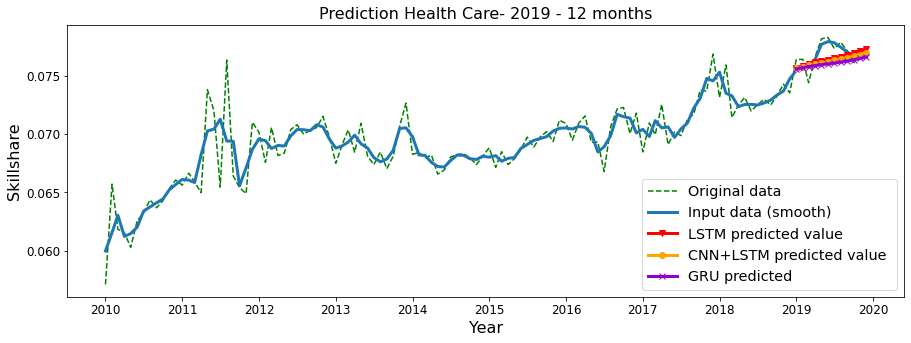}

    \end{tabular}
     \caption{Prediction results for business management and Effective Communication from Business skills dataset, Java and Python from Python dataset and Education and Training and Health care from Skill Family - Industry sectors dataset using LSTM, CNN+LSTM and GRU.}
    \label{fig:plot_multi_all_archi}
\end{figure*}

There is wide variation in the predictive ability of the models between skill datasets. In the multivariate experiment, predictions are best for the `Business management' skills with a NRMSE of 0.08 and worst for the `Cloud computing' skills dataset with a NRMSE of 0.42. `Business management' skills seem to appear more frequently in job postings than `Cloud computing', suggesting that demand for more frequently demanded skills might be less volatile and easier to predict. 

Table~\ref{tab:multi} depicts the performance of the three multivariate architectures for different time horizons (months ahead) forecasts in all the datasets built using the skill similarity method. Note that all these are one-shot predictions. We found that for all modeling approaches, the forecasting error (NRMSE) increases with the length of the forecasting horizon as expected. Figure \ref{fig:monhs_ahead} exhibits this upward trend of NRMSE as the horizon is increased from 6 months to 36 months. One exception to this trend is `Cloud computing' whose 12-months-ahead forecasts perform worse than 24-months-ahead output across all three architectures. 
For all variations of experiments, the one-shot 6-months-ahead prediction always yields the best result. The worst performance is recorded by the `Machine Learning' dataset for a 36 month prediction horizon. Predictions for the `Precision Agriculture' dataset were relatively bad for 6 months ahead; despite not being a new terminology, this dataset's frequency of mentions has only grown recently, indicating a lower frequency of mentions in the historical data.

 \begin{table}[!h]
  \caption{Skill Cluster Family (industry sectors) predictions 6, 12, 24 and 36 months ahead in a multivariate model}
    \label{tab:multi_skillclusterfamily}
 \centering
     \begin{tabular}{lcc}
\multicolumn{3}{c}{CNN+LSTM}\\	

Months ahead	&	NRMSE	&	MAPE	\\
\midrule
6	&	0.015	&	1.625	\\
12	&	0.024	&	2.189	\\
24	&	0.055	&	4.442	\\
36	&	0.085	&	6.692	\\
   \bottomrule
    \end{tabular}
   
\end{table}

`Skill Cluster Families', being EBG aggregations of several hundred skills and mapping roughly to career areas like Information Technology, Finance, etc., have smoothed skillshare time series because they are more aggregated. Table~\ref{tab:multi_skillclusterfamily} shows smaller errors (NRMSE) of prediction compared to the other datasets formed from similarity-based clusters of individual skills.
To be able to compare the performance of our approach with that of the ARIMA based approach in~\cite{das2020aies}, we compute and present the mean absolute percentage error (MAPE) of prediction of the Skill Cluster Families using the CNN+LSTM architecture. 
The cited article reports an average MAPE of 1.38\% for one-month-ahead forecasts on the Skill Cluster Families dataset. This work can predict six months ahead achieving 1.63\% of MAPE. Note that the performance of deep learning-based architectures proposed in this paper are much better than the ARIMA as it can only forecast one month ahead. To further demonstrate the advantages of our framework over other state-of-the-art skills demand forecasting approaches, we compare with the work of Das \emph{et al}~\cite{das2020aisg} where the 12-month and 24-month ahead predictions by the same authors yield about 10\% MAPE, whereas our approach report only 2.2\% and 4.4\% MAPE for those two time horizon forecasts.

\begin{figure}[!t]
    \centering
    \begin{tabular}{c}
      \includegraphics[width=1\linewidth]{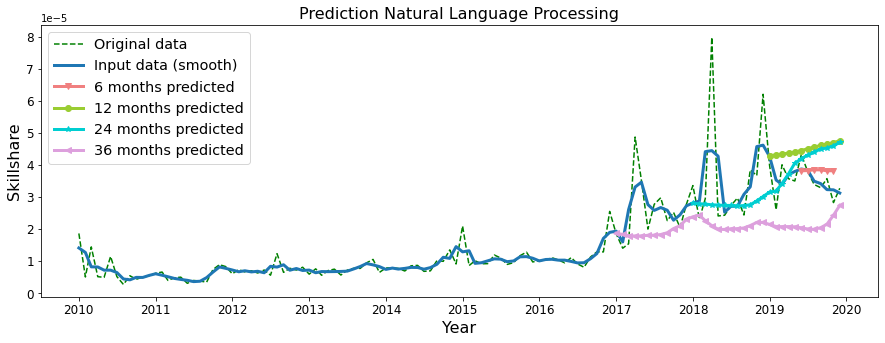}   \\
      \includegraphics[width=1\linewidth]{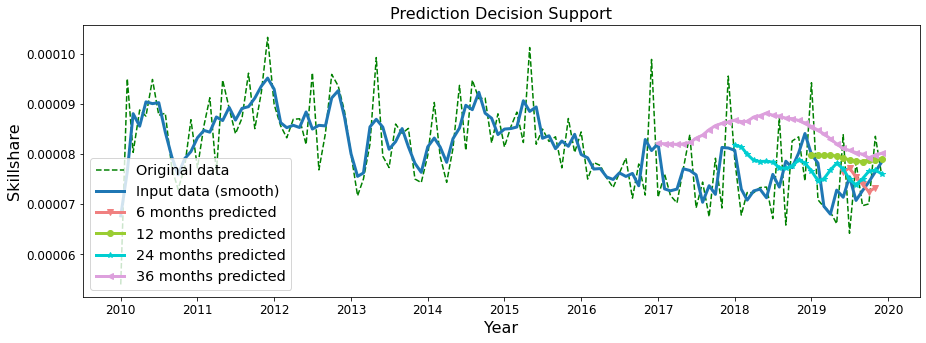} \\  
    \end{tabular}
     \caption{Prediction results along months ahead to predict for Natural Language processing from Quantum dataset and Decision Support from the dataset with the same name.}
    \label{fig:months_ahead_pred}
\end{figure}

For visual illustration of the dataset and forecasting, we present Figure ~\ref{fig:plot_multi_all_archi}, showing the prediction performance of a few selected skills across all three architectures for the multivariate problem. We display the individual skills `Business management' and  `Effective communication' from the `Business management' dataset, skill `Python' and `Java' from the `Python' dataset, and finally skills `Education and training' and `Health care' from the `Industry sectors' dataset. In these examples, 12-months-ahead predictions of all the skills, excepting `Business management', were able to track the trends of the actual time-series. Another way to visually illustrate results of the multivariate experiment is by comparing forecasts for different time horizons, as presented in Figure~\ref{fig:months_ahead_pred}.

\begin{figure}[!t]
    \centering
    \begin{tabular}{c}
      \includegraphics[width=1\linewidth]{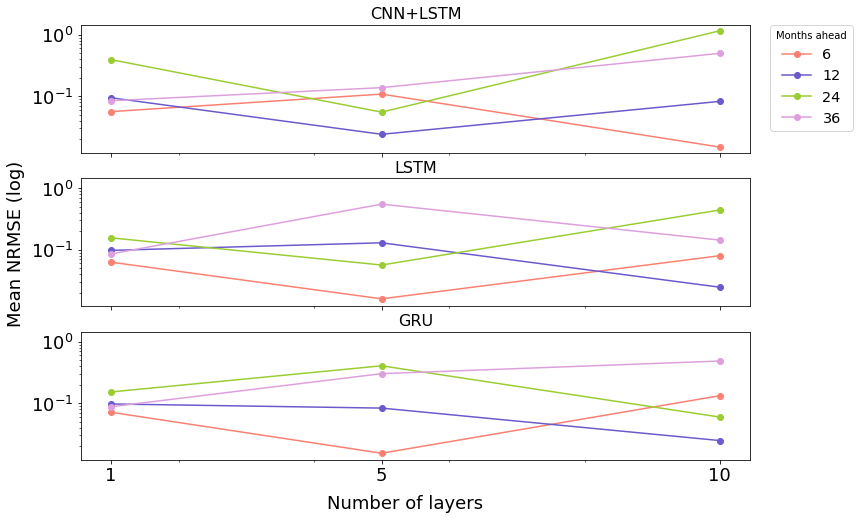}    
    \end{tabular}
     \caption{Analysis of performance on increasing number of layers for LSTM and GRU architectures.}
    \label{fig:plot_layers}
\end{figure}

While carrying out extensive experiments varying many hyperparameters, we also tried varying the number of layers in the recurrent network architectures. Figure~\ref{fig:plot_layers} shows that there is no evidence that increasing the number of layers can further improve the results. This stands to reason since the skillshare timeseries data are short longitudinally. 120 time-points of data (months) is not enough to leverage the benefits of very deep architectures. As we have one optimized multivariate model for each dataset, Table~\ref{tab:hyperparameter} depicts the hyperparameters which return the best performance in one-shot 12-months-ahead prediction.

\begin{figure}[!t]
     \centering
     \begin{tabular}{c}
       \includegraphics[width=1\linewidth]{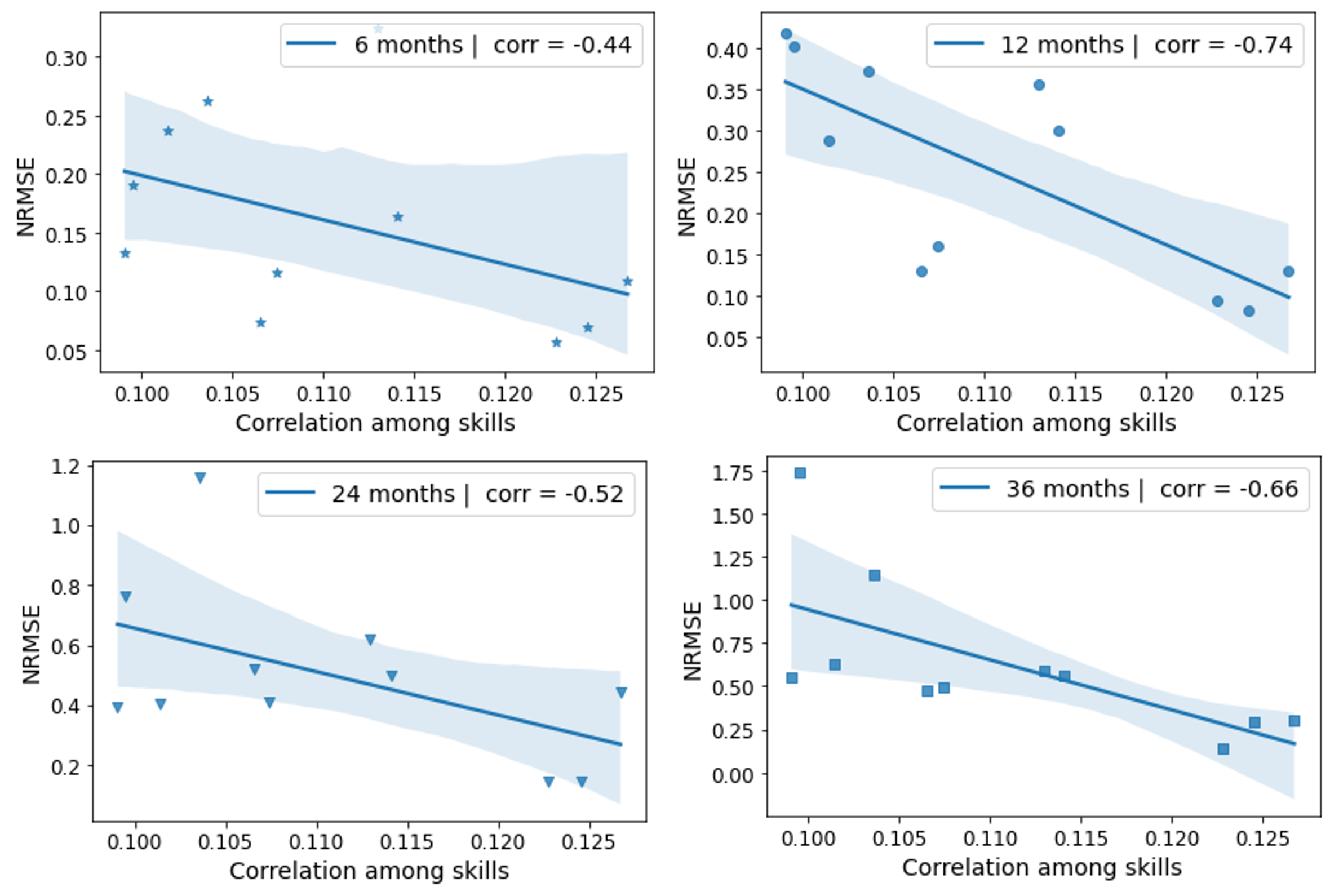}   \\  
     \end{tabular}
     \caption{Minimum skills correlation versus NRMSE. Minimum correlation means the smallest 
     correlation value between \textit{n} skills in a cluster.}
     \label{fig:plot_correlation}
 \end{figure}

Our rationale behind the potential strength of a multi-variate approach was that it may yield better results by learning from coupling (correlation) among skills in the datasets.
Our hypothesis is that in the job market the demand for some skills is dependent on the demand for others, directly or even inversely. Figure~\ref{fig:plot_correlation} supports this hypothesis, showing that NRMSE has a downward trend (negative correlation) with respect to coupling (correlation) among skills across all eleven skills datasets. The coupling is measured by minimum correlation, i.e., the smallest correlation value between \textit{n} skill pairs within a cluster. 

In an attempt to introduce and establish a practical tool for skill demand forecasting, we chose well known state-of-the-art RNNs, keeping in mind limitations regarding the length of longitudinal data required for training deeper neural network architectures. In future work, we plan to explore a transformer-based approach, which is a method based on the Multihead-Self-Attention (MSA) mechanism, to experiment on the forecasting of skill datasets. According to Khan \emph{et al}~\cite{Khan}, transformers enable long dependencies between input sequence elements, support parallel processing of sequences as compared to RNN, and demonstrate excellent scalability to very large capacity networks and huge data. Given that we focused on computer-related occupations and IT skills as sample subsets of the full skills demand data, we also leave extensions and generalizations of the proposed pipeline to other occupations and labor market skills as future work. 

Since Molloy~\emph{et al}~\cite{Molloy2017} affirm that interstate migration in the United States has decreased steadily since the 1980s, one of these future extensions can be the consideration of states, locales or commuting zones in our methodology, thus bringing more specific information to interested parties.

\begin{table*}[]
\caption{Hyperparameters of a multivariate model for the best performance considering the one-shot 12 months ahead prediction}
\label{tab:hyperparameter}
\centering
\begin{tabular}{lcccccc}

DATASET	&	model	&	\#layers	&	\# neurons	&	kernel	&	epochs	&	lag	\\
  \midrule
Agile Development	&	CNN+LSTM	&	5	&	10	&	16	&	100	&	12	\\
Business Management 	&	CNN+LSTM	&	10	&	10	&	64	&	2000	&	36	\\
Cloud Computing     	&	CNN+LSTM	&	1	&	2	&	64	&	500	&	12	\\
Cryptography        	&	GRU	&	1	&	10	&	---	&	100	&	36	\\
Data Science       	&	CNN+LSTM	&	5	&	10	&	64	&	100	&	36	\\
Decision Support   	&	CNN+LSTM	&	1	&	2	&	16	&	500	&	36	\\
Machine Learning   	&	CNN+LSTM	&	10	&	10	&	2	&	500	&	12	\\
Precision Agriculture 	&	CNN+LSTM	&	1	&	10	&	2	&	50	&	12	\\
Python 	&	LSTM	&	5	&	10	&	---	&	50	&	36	\\
Quantum Computing   	&	CNN+LSTM	&	1	&	1	&	64	&	500	&	36	\\
Security Assessments	&	LSTM	&	1	&	2	&	---	&	100	&	24	\\
Skill C. Family- Industry Sectors	&	CNN+LSTM	&	5	&	2	&	2	&	50	&	36	\\

\bottomrule
    \end{tabular}
\end{table*}

\subsection{Use Case and Impact}
\label{sec:impact}

The pipeline presented in this paper could be useful for any organization that currently utilizes up-to-date skills data as an implicit forecast. Many community colleges and universities subscribe to labor market data providers such as EBG, LinkedIn Talent Insights\footnote{\url{https://business.linkedin.com/talent-solutions/talent-insights}}, and Gartner TalentNeuron\footnote{\url{https://www.gartner.com/en/human-resources/research/talentneuron}} to inform their course selections and curricula. Subscribers might benefit if a forecasting capability were added to the data providers' offerings.

One specific organization that could utilize a skills forecasting pipeline is the National Center for O*NET Development, which develops and maintains the Occupational Information Network (O*NET), a U.S. Department of Labor-sponsored database of occupation profiles. O*NET maintains lists of skills commonly utilized in each occupation but has been criticized for slow incorporation of new skills and technologies into their ontology \cite{Tippins2010}. O*NET already works with EBG to monitor job posting data for recent changes in skill demand within occupations \cite{BurningGlass2019}; forecasts like those demonstrated in this paper could provide earlier warning \emph{before} skill demand crosses a relevant threshold.

Likewise, employers with work portfolios in areas of rapidly evolving technology can benefit from a skills forecasting pipeline. To maintain a relevant workforce, such businesses often face a choice of whether to lay off existing workers and hire new people or reskill their existing employees. Laying off workers is difficult for those losing their job and bad for the morale of employees who remain, and hiring new workers is expensive. But reskilling requires time, which companies often fear they cannot afford. Forecasts might buy time for reskilling, improving workers' welfare and increasing labor market efficiency by preserving the value of proven employer-employee matches.

It should be emphasized that political or economic shocks will not enter into our model until they are reflected in firms' advertising behavior. However, since job ad data are available in real time, our model should pick up new trends relatively quickly. For example, changes to skill demand resulting from a government initiative to build domestic chip manufacturing capacity would not be forecast until the hiring trend has already started. But once a change in advertising behavior begins, our model would quickly pick it up.

\section{Conclusions}
\label{sec:conclusion}

In  this  paper, we apply three recurrent neural networks to a skill demand forecasting task,  utilizing data from  January 2010  until  December  2019.   We address the idea of performing a multivariate prediction for a large number of skills and compare it to separate univariate predictions for each skill. Although the results of the univariate model are better than the multivariate in the case where we independently searched for the best model for each skill, it does not necessarily mean that the solution would be to choose univariate models for prediction. Given that many skill taxonomies contain thousands of skills, the training and re-training of many models could become unfeasible. 
Furthermore, results of the multivariate model still provide a good indication of the trend of the time series for the next full year. We believe this exercise confirms the usefulness of both multivariate and univariate skills demand forecasting for mitigating the skills gap and helping students and workers to pursue their professional pathways. 

\bibliographystyle{unsrtnat}
\bibliography{main}

\end{document}